# In-line extraction of an ultra-stable frequency signal over an optical fiber link

**Anthony Bercy,[1,2] Saïda Guellati-Khelifa,[3] Fabio Stefani,[2,1] Giorgio Santarelli,[4] Christian Chardonnet,[1] Paul-Eric Pottie,[2] Olivier Lopez,[1] and Anne Amy-Klein[1,*]**

*[1]Laboratoire de Physique des Lasers, Université Paris 13, Sorbonne Paris Cité, CNRS, 99 Avenue Jean-Baptiste Clément, 93430 Villetaneuse, France*

*[2]Laboratoire National de Métrologie et d'Essais–Système de Références Temps-Espace, UMR 8630 Observatoire de Paris, CNRS, UPMC, 61 Avenue de l'Observatoire, 75014 Paris, France*

*[3]Laboratoire Kastler-Brossel, UPMC, ENS, CNRS, 4 place Jussieu, case 74, 75005 Paris, France*

*[4]Laboratoire Photonique, Numérique et Nanosciences, Université de Bordeaux 1, Institut d'Optique and CNRS, 351 cours de la Libération, 33405 Talence, France*

**Corresponding author: amy@univ-paris13.fr*

We demonstrate in-line extraction of an ultra-stable frequency signal over an optical link of 92-km of installed telecommunication fibers, following the proposition of G. Grosche in 2010 [1]. We show that the residual frequency noise at the extraction end is noticeably below that at the main link output when the extraction is near the input end, as expected from a simple model of the noise compensation. We obtain relative frequency instabilities, expressed as overlapping Allan deviation, of $8\times10^{-16}$ at 1 s averaging time and a few $10^{-19}$ at 1 day. These results are at the state-of-the-art for a link using urban telecommunication fibers. We also propose an improved scheme which delivers an ultra-stable signal of higher power, in order to feed a secondary link. In-line extraction opens the way to a broad distribution of an ultra-stable frequency reference, enabling a wide range of applications beyond metrology.

## 1. INTRODUCTION

For a decade, optical fiber links have experienced a lot of developments all around the world and it is now established that they are a very efficient technique for ultra-stable frequency transfer on a continental scale. The better performance has been obtained by transferring an ultra-stable optical frequency around 1.55 µm [2]. Optical frequency transfer has been demonstrated by several groups on distances up to a few hundreds of km [2-6] and used for clocks frequency comparison [7-10]. Record optical links of 920 km and 1840 km were demonstrated in Germany with a frequency transfer stability (modified Allan deviation) of $3\text{-}5\times10^{-15}$ at 1 s averaging time and below $10^{-18}$ after a few hundred of s [11, 12]. These links use optical fibers dedicated to this application. Our groups demonstrated an optical link of 540 km using telecommunications fibers, one frequency channel being devoted to ultra-stable frequency transfer, the other channels carrying simultaneously Internet data traffic [13]. The stability performances are similar to the ones obtained with dedicated fibers. Such a "dark-channel" approach opens the possibility to develop a broader fiber network. This is an issue not only for clock comparisons in national metrology institutes but also for high-resolution spectroscopy and/or remote laser stabilization performed in many research laboratories, which may benefit from an ultra-stable frequency reference. That is the purpose of the REFIMEVE+ project, a national metrological network connecting more than 20 research laboratories in France.

But the point-to-point transfer scheme, as demonstrated up-to-now, requires a fiber link per user. This is not suitable for distributing the ultra-stable frequency signal to many users, especially in a metropolitan area network, since it would need a lot of fibers. As proposed by G. Grosche three years ago [1], it is more efficient to implement a main optical link along which the signal is distributed to multiple users. Another possibility is to implement a branching optical fiber network with noise correction at each output end [14]. In this paper, following first implementations on a few-km fiber spools [15-17], we demonstrate for the first time a multiple-access frequency dissemination using installed telecommunication fibers of 92 km. Moreover we show that the residual frequency noise at the extraction end is below that at the main link output. The paper is organized as follows. In a first part, we describe the experimental set-up which consists in extracting the signal from a main link to a local user. We also give a simple estimation of the residual noise at the extraction end. Then we show experimental results using dedicated telecommunication fibers for an extraction point near the input or the output end of the main link. Finally we propose an improved extraction scheme in order to distribute the ultra-stable signal with a longer range for the distant user using a secondary link.

## 2. EXPERIMENTAL SCHEME

Figure 1 shows the scheme of the multiple-access frequency transfer with a simple extraction along the main link, as proposed and demonstrated on fiber spools by G. Grosche [1, 17]. The main optical link aims at copying the ultra-stable signal of the input end of the fiber to the output end. But the signal phase is disturbed by the fiber noise arising from temperature fluctuation and mechanical vibration. The challenge is therefore to compensate these disturbances at the output end and at any point along the fiber, in order to distribute the ultra-stable signal to several users at once. In this paper, we will use the following notations: the transferred laser signal has a total phase $2\pi\nu t + \phi(t)$ where $\nu$ is the laser frequency and $\phi(t)$ the phase perturbation.

The main link is composed of two sections of optical fibers of lengths $L_A$ and $L_B$ respectively, with total length $L = L_A + L_B$, and two stations at each end. In the input station, the ultra-stable reference signal is provided by a laser emitting at 1.55 µm, stabilized to an ultra-stable Fabry-Perot cavity. Its frequency $\nu_0$ is eventually measured or controlled with primary frequency standards at SYRTE using an optical frequency comb [18, 19]. The laser stability is around $10^{-15}$ at 1 s averaging time ensuring that the laser noise does not impact the frequency transfer [5]. The laser frequency $\nu_0$ is up-shifted by frequency $f_1$ with an acousto-optic modulator (AOM) denoted as AOM1 and then feeds the fiber. At the output end, it is again up-shifted by frequency $f_2$ with AOM2, and part of it is sent back with a Faraday mirror. The round trip signal is mixed with the input ultra-stable laser using an interferometer made of an optical coupler (OC) and a Faraday mirror. The beat-note frequency is twice the sum of the AOMs frequencies. For clarity, we first neglect the propagation delays along the fiber. In that case, the beat-note signal exhibits the round-trip fiber phase noise, $2\phi_F$, with $\phi_F = \phi_A + \phi_B$, where $\phi_A$ and $\phi_B$ are the noise of the fiber sections of length $L_A$ and $L_B$ respectively. After down conversion to 0-frequency and signal processing, the phase correction $\phi_{C1}$ is applied on the AOM1 frequency driver. Since this correction is applied both on the forward and backward signal, we have $\phi_{C1} = -\phi_F$. Finally at output end, the phase fluctuations are $\phi_{C1} + \phi_F$ and thus cancel.

We now consider the extraction of the ultra-stable signal along the fiber, at a distance $L_A$ from the input end and $L_B$ from the output end (Fig. 1). An OC enables to extract both the forward and backward signals from the fiber. The forward signal has a frequency $f_1$ and exhibits the phase fluctuations $\phi_{C1} + \phi_A = -\phi_B$. This shows that the main link compensation loop induces an over-correction of the fiber phase noise on the forward extracted signal. To compensate it, we detect the beat-note of the two extracted signals [17]. A polarization controller is inserted in one arm of the Mach-Zehnder interferometer in order to align the polarizations of the two extracted beam, which are perpendicular at the output of coupler OC1. The backward signal has a frequency $f_1 + 2f_2$ and exhibits the phase fluctuation $\phi_{C1} + \phi_A + 2\phi_B = \phi_B$. The beat-note frequency is thus $2f_2$ and it exhibits the phase fluctuation $2\phi_B$. The signal frequency is divided by 2, filtered and drives an AOM (AOM3) in order to correct for the frequency and phase fluctuation of the forward extracted signal. The frequency of the forward extracted signal, after passing through AOM3, is thus up-shifted to $\nu_0 + f_1 + f_2$ and its phase fluctuation cancelled. Similar compensation can be obtained on the backward extracted signal with a negative frequency shifter. This set-up also compensates any frequency fluctuation of the AOM2 at the output end, which may be caused by the fluctuation of the distant RF oscillator. Therefore no ultra-stable RF oscillator is required for this set-up, except at the input end.

Although very simple, this set-up is sensitive to the fiber noise of the extraction set-up, due to uncompensated fiber paths [20]. An improved extraction design, so-called "tentacle", has been proposed in [17] and enables to be insensitive to the noise of the fiber sections at the output of the extraction coupler OC1. We chose to keep the simple design of Fig. 1 but we carefully set the fiber lengths in order to minimize uncompensated fiber paths. The contribution of the local fiber noise at the output of the extraction set-up is given by $\phi_{lf} = \phi_{L1} + \phi_{L4} + (\phi_{L2} - \phi_{L3})/2$ where $\phi_{Li}$ is the noise arising from the fiber section of length $L_i$ (see Fig. 1). We set these fiber lengths such that the noise $\phi_{lf}$ cancelled: in the case of an homogeneous temperature, this simply requires that $L_3 = L_2 + 2(L_1 + L_4)$. With that scheme, we unbalance the fiber paths of the two extracted signals at the optical coupler OC2 in order to detect a fiber noise equivalent to a length $L_1+L_4$ and compensate for it with the feed-forward correction applied through AOM3.

The above description does not take into account propagation delays inside the fiber sections. It was first pointed out by Newbury and coworkers [3, 21] that it limits the noise rejection proportionally to the square of the propagation delay. In Appendix A we detail the calculation of the noise compensation for both the main link output and the extracted signal for the set-up of Fig. 1. In the case of an uncorrelated and position-independent fiber noise along the two fiber sections, we find that the phase noise power spectral density (PSD) of the extracted signal, $S_E(f)$, is related to the phase noise PSD at the link output, $S_O(f)$, as follows : $S_E(f) = F\, S_O(f)$

$$\text{with } F = \left(\frac{L_A}{L}\right)^2 \left(3 - 2\frac{L_A}{L}\right)$$

This factor F is null for $L_A=0$, then increases up to 1 at $L_A=L$. With that model, we see that the extraction phase noise PSD is at most equal to the output phase noise PSD and can be significantly below for $L_A \ll L$. This behavior can be explained as follows. When the extraction occurs at the link input ($L_A=0$), the phase correction at the extraction is exactly opposite to the roundtrip fiber phase noise and thus to the correction at the link input (see Appendix A). Therefore the extracted signal is exactly the input signal and no fiber phase fluctuation is added to the signal. Along the link, this compensation is no more valid and the residual noise increases. Thus, near the link input ($L_A \ll L$), a major part of the residual noise is compensated. Near the link output the noise of the second fiber section $L_B$ is negligible and the forward extracted signal is approximately the same than the output signal.

## 3. EXPERIMENTAL RESULTS

We implement the extraction scheme of Fig. 1 on a 92-km optical link using installed fibers of the telecommunication network (Fig. 2). The main link input is at SYRTE and is composed of two sections of 86 km and 6 km. The first one is using two parallel dedicated fibers of 43 km linking SYRTE to LPL [5], the second one is using two parallel dedicated fibers of 3 km linking SYRTE to LKB [22]. That way, the two ends of the main link and the extraction set-up are at the same place, at SYRTE. This configuration gives the possibility to detect the beat-notes between the link ends and characterize the noise compensation. AOM1 and AOM2 frequencies are respectively 39 MHz and 37 MHz, therefore both the main link end-to-end beat-note's and the beat-note between the input end and the extracted end have a frequency of 76 MHz. These beat-notes are tracked with a 100 kHz bandwidth and after frequency division by 76 they are recorded simultaneously with two dead-time free counters (Kramer+Klische FXE) with a gate time of 1 s and Π-type and Λ-type operation respectively. These two types of counting give complementary information on the stability of the extracted signal. The Π-type samples give the possibility to calculate directly the Allan deviation stability of the extracted signal. With the Λ-type data, noise is averaged out in a narrower bandwidth and we can probe both the various phase noise dependence and the noise floor. Detailed discussions on the properties of overlapping and modified Allan deviation can be found in [23, 24].

Fig. 3a displays the experimental stability (green stars) obtained when the extraction is done after the 86-km fiber section (Fig. 2). It is calculated using overlapping Allan deviation and Π-type counting. The extraction stability is $1.3 \times 10^{-15}$ at 1 s averaging time, decreases as expected with a $\tau^{-1}$ slope and reaches a floor of $10^{-18}$ after 2000 s. It is at the state of the art for fiber frequency transfer on such distances [5]. A better stability has been obtained with a similar extraction set-up [16, 17] but with fiber spools of up to 3 km, which are much less noisy. The extraction end stability is very similar to the end-to-end stability (blue diamonds), except for averaging time longer than 2000 s. In this configuration, the F factor introduced above is equal to 0.99, thus we expect that the extraction end stability copies the output stability. We obtained an experimental value F=0.96, compatible with the model. For longer averaging time, the extraction end stability is limited by the noise floor, which is about $10^{-18}$ between $10^3$ and $2 \times 10^4$ s. This noise floor arises mainly from the local fiber noise at the extraction end, which exhibits still some uncompensated fiber paths. The fiber lengths were not enough precisely adjusted such that the noise cancelled in the case of a homogeneous temperature (see fourth paragraph of section 2). Moreover we did not implement an active temperature stabilization. This induces a local fiber noise due to temperature residual fluctuations. It is in fact a preliminary set-up for testing the compensation method. By contrast, the end-to-end noise floor is below $3 \times 10^{-19}$ after 1000 s and does not limit the output end stability since the main link stabilization and measurement set-ups have been carefully optimized following few years of development [13, 25]. Fig. 3a displays also the end-to-end free-running stability (red triangles) and the extraction-to-end stability (black squares) when the main link is compensated but the extraction correction is not applied to AOM3. We clearly see that the phase noise at the extraction end is over corrected and exhibits a stability of $10^{-14}$ at 1s, which is one order of magnitude higher than the end-to-end stability. We also plot on Fig. 3a the modified Allan deviation at the output extraction end calculated from Λ-type data (dark green circles). It is $2 \times 10^{-17}$ at 1 s averaging time and reaches around $4 \times 10^{-19}$ after $10^3$ s. As explained above, although very low, this stability is limited by the noise floor due to the temperature fluctuation at the extraction end, which induces fluctuations of the optical phase.

Fig. 3b displays the experimental stability (Π-type counting, overlapping Allan deviation) (green stars) obtained when the extraction is done after the 6-km fiber section. The extraction stability is $8 \times 10^{-16}$ at 1 s averaging time, decreases with a $\tau^{-1}$ slope up to a few 100 s and reaches a floor of $5 \times 10^{-19}$ at $10^4$ s. For this long time, the extraction stability is limited by temperature effects on uncompensated fiber paths as highlighted above. The extraction stability is better than the end-to-end stability (blue diamonds) up to an averaging time of 1000 s. In this configuration, the F factor introduced above is equal to 0.01, thus we expect that the extraction end stability is significantly below the output stability. Here the experimental value of F is about 0.58. This discrepancy can be explained by the fact that the fiber noise is far from being position-independent along the fibers, since they are in a metropolitan area. Moreover in the model of Appendix A we do a first-order development of the phase-noise versus time, and thus we neglect the second time-derivative of the phase noise. Fig. 3b displays also the output end free-running stability (red triangles) and the extraction end stability (black squares) when the main link is compensated but the extraction correction is off. The phase noise at the extraction end is quite similar to the free-running noise of the main link, which is expected for an extraction near the input end, where the over-correction is maximum. Finally we also plot on Fig. 3b the modified Allan deviation at the output extraction end calculated with Λ-type data (dark green circles). It is $1.1 \times 10^{-17}$ at 1 s averaging time and reaches around $4 \times 10^{-19}$ after 2000 s and is still limited by the noise floor.

We also performed an evaluation of the accuracy of the frequency transfer at the extraction end. We first calculated the mean frequency of the end-to-extraction beat-note's data, recorded with 1 s gate time and Π-type counter, over successive 100 s segments. Then we calculated the mean value and the standard deviation of these frequencies (weighted with the inverse of the variance of each 100 s sample). Fig. 4 displays the data for the extraction after 6 km (4786 samples of 100 s). The mean frequency is shifted by $-1.1 \times 10^{-19}$, with a standard deviation of $6.2 \times 10^{-18}$. For the extraction after 86 km (404 samples of 100 s), the mean frequency is shifted from the expected value $f_1 + f_2$ by a fractional value of $-2.4 \times 10^{-18}$, with a standard deviation of $1.1 \times 10^{-17}$. The statistical uncertainty on the fractional mean frequency is therefore $9 \times 10^{-20}$ and $5.3 \times 10^{-19}$ respectively for the near or far extraction and this demonstrates that there is no systematic frequency shift arising in the extraction set-up at a level of a few $10^{-19}$.

## 4. IMPROVED EXTRACTION SCHEME

The above extraction scheme, as highlight in [17], is very simple and efficient. Provided the main link is lengthened and the extraction coupler OC1 is put near the user set-up, it can provide an ultra-stable signal to any user close to the back-bone main link. However the extracted signal power may be too low for direct applications, when the main link is long enough and exhibits significant losses. In that case, we propose to use a narrow bandwidth laser diode for optical regeneration and amplification. Our proposal is sketched on Fig. 5. A laser diode is offset phase-locked to the forward extracted signal and feeds a secondary optical link in order to distribute the extracted signal to a distant laboratory. The frequency offset between the laser diode and the extracted signal is equal to the sum of the local oscillator frequency $f_{LO}$ and the frequency $f_2$, that is half the beat-note frequency between the forward and backward extracted beams. The laser diode frequency is thus $\nu_{DL} = \nu_0 + f_1 + f_2 + f_{LO}$ and its phase fluctuations are copying the phase fluctuation $\phi_{LO}$ of the local oscillator (within the loop bandwidth). The noise of the secondary optical link $\phi_S$ is compensated with the usual round-trip method. Corrections are applied through a negative frequency shift AOM, denoted AOM3, which induces a frequency shift of $-f_{LO}$, and a phase correction $-\phi_{C2} = -(\phi_{LO} + \phi_S)$. Thus the secondary link output frequency is copying the ultra-stable frequency at the main link output $\nu_0 + f_1 + f_2$ without any added phase noise. Although using a local RF oscillator at the extraction end, this set-up is insensitive to its frequency fluctuations. More details are displayed on Fig. 5, where the phase perturbations are indicated. A second extraction output can be added to this set-up at the laser diode output, provided that a negative-shift AOM is added to compensate the local oscillator frequency fluctuations. This second output enables to distribute the ultra-stable signal to a second user or eventually many users.

Following the same approach as for the first set-up, we designed this set-up to minimize the noise resulting from uncompensated fiber paths. We chose to make identical the fiber lengths ($L_1+L_2$) of both fiber arms between the optical couplers OC1 and OC2. We also set the fiber length between the laser diode and the optical coupler OC4 to be the sum of the fiber lengths between OC1 and OC4 ($L_1+L_2$) and between the laser diode output and the Faraday mirror after OC6 ($L_3+L_4$) (see Fig. 5). That way, we add to the laser diode phase the opposite of the uncompensated fiber noise between the laser output and the Faraday mirror near OC6.

Although more complex, this set-up is currently under development and will be installed in the center of Paris in order to distribute the signal to several laboratories, including LKB. A secondary link is necessary in order to reach the experimental room of LKB which is far enough (a few tens of m) from the extraction point. Moreover this scheme will also give the possibility in the future to distribute the ultra-stable signal to laboratories about 20 km to the south of Paris.

## 5. CONCLUSION

In summary, we have demonstrated the distribution of an ultrastable frequency at an intermediate extraction point along an optical fiber link. It was demonstrated with two fiber sections from the installed telecommunication network of 6 and 86 km respectively. The ultra-stable signal extraction was demonstrated with stability (modified Allan deviation, Λ-type counting) of $1.1 \times 10^{-17}$ at 1 s averaging time and below $10^{-18}$ after 500 s. This stability is at the state of the art of optical links and can be further improved at long term with better minimization of the uncompensated fiber paths and better temperature stabilization. The residual frequency noise at the extraction end is at least as low as the noise at the main link output, in agreement with a simple model. An improved extraction design has been proposed which gives the possibility to obtain an ultra-stable signal of higher power, and is insensitive to the frequency fluctuations of the RF local oscillator. Such an extracted signal can feed a secondary optical link for distribution on a metropolitan area scale.

This extraction set-up opens the way to a broad distribution of an ultra-stable frequency reference, enabling applications beyond metrology. It is a key-element of the current effort to establish national or even continental ultra-stable fiber network. In France, such an extraction set-up will be utilized for metropolitan area distribution, as for example in the Paris area. Specifically, it will be installed at Kastler-Brossel laboratory (LKB), located in the center of Paris, which already receives an ultra-stable RF signal from SYRTE [22] for high-resolution spectroscopy of Hydrogen [26] and high precision determination of the fine structure constant [27]. These experiments will benefit from the improved stability and accuracy of an optical frequency reference. We expect that such new development of ultra-stable frequency distribution will stimulate the development of new high-sensitivity experiments in a broad range of applications.

## ACKNOWLEDGMENTS

We are very grateful to François Biraben, François Nez and Pierre Cladé from LKB for fruitful discussions as well as giving us the opportunity to perform this experiment. We acknowledge financial support from the Agence Nationale de la Recherche (ANR BLANC 2011-BS04-009-01), Labex First-TF (ANR-10-LABX-48-01), the European Metrology Research Programme (EMRP) under contract

SIB-02 NEAT-FT and IFRAF-Conseil Régional Ile-de-France.

## APPENDIX A

In this appendix we analyze the noise rejection at the extraction output for the scheme of Fig. 1. Following the approach of [21], the phase noise accumulated by the optical carrier propagating along the main link in the forward direction and exiting the fiber at output end at time t is

$$\phi_{f+}(t)=\int_0^L \delta\varphi_z\left(t-\left(\tau-z/v\right)\right)dz$$

where z is the position coordinate along the fiber, with z=0 at input end and z=L at output end, $\tau=L/v$ is the end-to-end propagation delay in the fiber, $v=c/n$ is the speed of light in the fiber and $\delta\varphi_z(t)$ is the fiber phase noise per unit of length at position z and time t.

Similarly, the phase noise accumulated by the optical carrier propagating in the backward direction and exiting the fiber at input end at time t is

$$\phi_{f-}(t)=\int_0^L \delta\varphi_z\left(t-z/v\right)dz$$

Thus the roundtrip phase noise at time t is given by:

$$\phi_{f,RT}(t)=\phi_{f-}(t)+\phi_{f+}(t-\tau)$$

$$=\int_0^L \delta\varphi_z\left(t-z/v\right)dz+\int_0^L \delta\varphi_z\left(t-\left(2\tau-z/v\right)\right)dz$$

In the limit of an ideal correction loop, the correction phase is generated such as:

$$\phi_{f,RT}(t)+\phi_c(t)+\phi_c(t-2\tau)=0.$$

A first-order Taylor development gives:

$\phi_c(t)+\phi_c(t-2\tau)=2\,\phi_c(t-\tau)$ and

$$\phi_{f,RT}(t)=2\int_0^L\left(\delta\varphi_z(t)-\tau\,\delta\varphi_z'(t)\right)dz$$

where $\delta\varphi_z'(t)$ is the time derivative of $\delta\varphi_z(t)$

Thus $\phi_c(t-\tau)=-\int_0^L\left(\delta\varphi_z(t)-\tau\,\delta\varphi_z'(t)\right)dz$.

We now calculate the residual phase noise power spectral density (PSD) at the main link output.

The phase noise at the output end at time t is:

$$\phi_o(t)=\phi_{f+}(t)+\phi_c(t-\tau)$$

At first order, it simplifies to:

$$\phi_o(t)=\int_0^L\left(\frac{z}{v}\,\delta\varphi_z'(t)\right)dz$$

We calculate the autocorrelation function of $\phi_o(t)$.

$$R_o(u)=\overline{\phi_o(t)\phi_o(t-u)}=\int_0^L\int_0^L\left(\frac{z}{v}\frac{z'}{v}\,\overline{\delta\varphi_z'(t)\delta\varphi_{z'}'(t-u)}\right)dz\,dz'$$

We assume that the fiber noise time derivative is not correlated in position:

$\overline{\delta\varphi_z'(t)\delta\varphi_{z'}'(t-u)}=\overline{\delta\varphi_z'(t)\delta\varphi_z'(t-u)}\,\Delta L\,\delta(z-z')$ with $\Delta L$ the width of the spatial autocorrelation function [28], and that $\overline{\delta\varphi_z'(t)\delta\varphi_z'(t-u)}$ is independent of z.

Then

$$R_o(u)=\int_0^L\left(\frac{z}{v}\right)^2\overline{\delta\varphi_z'(t)\delta\varphi_{z'}'(t-u)}\,\Delta L\,dz$$

$$=\overline{\delta\varphi'(t)\delta\varphi'(t-u)}\,\Delta L\,\frac{L^3}{3v^2}$$

Finally we deduce the phase noise PSD at the output end as the Fourier transform of $R_o(u)$ :

$$S_o(f)=\mathcal{F}\left(R_o(u)\right)=\frac{\tau^2}{3}\mathcal{F}\left(\overline{\delta\varphi'(t)\delta\varphi'(t-u)}\,\Delta L\,L\right)$$

where $\mathcal{F}(\ )$ denotes the Fourier transform and $\mathcal{F}\left(\overline{\delta\varphi'(t)\delta\varphi'(t-u)}\,\Delta L\,L\right)=(2\pi f)^2\,\mathcal{F}\left(\overline{\delta\varphi(t)\delta\varphi(t-u)}\,\Delta L\,L\right)$ is approximatly the fiber phase noise derivative PSD. This formula is similar to the one derived in [21].

We now calculate the residual phase noise power spectral density (PSD) at the extraction output.

The phase noise accumulated by the optical carrier propagating along the fiber section $L_A$ in the forward direction and exiting the fiber at time t is

$$\phi_{fA+}(t)=\int_0^{L_A}\delta\varphi_z\left(t-\left(\tau_A-z/v\right)\right)dz$$

where $\tau_A=L_A/v$ is the propagation delay in the first fiber section.

Similarly, $\phi_{fB+}(t)$ and $\phi_{fB-}(t)$ are the phase noises accumulated by the optical carrier propagating along the fiber section $L_B$ in the forward and backward direction respectively and exiting the fiber at time t:

$$\phi_{fB+}(t)=\int_{L_A}^{L}\delta\varphi_z\left(t-\left(\tau-z/v\right)\right)dz$$

$$\phi_{fB-}(t)=\int_{L_A}^{L}\delta\varphi_z\left(t+\tau_A-z/v\right)dz.$$

The phase noise at the extraction output end at time t is thus:

$$\phi_e(t)=\phi_{fA+}(t)+\phi_c(t-\tau_A)+\frac{1}{2}\phi_{PD}(t)$$

where $\phi_{PD}(t)$ is the beat-note phase detected with the photodiode :

$$\begin{aligned}\phi_{PD}(t)&=\phi_c(t-\tau_B-\tau)+\phi_{fA+}(t-2\tau_B)+\phi_{fB+}(t-\tau_B)\\&\quad+\phi_{fB-}(t)-\phi_c(t-\tau_A)-\phi_{fA+}(t)\end{aligned}$$

with $\tau_B = {}^{L_B}\!/_{v}$ the propagation delay in the fiber section $L_B$..

At first order approximation, it gives:

$$\begin{aligned}\phi_e(t)&=\phi_c(t-\tau)+\int_{L_A}^{L}\left(\delta\varphi_z(t)-\tau_B\delta\varphi'_z(t)\right)dz\\&\quad+\int_{0}^{L_A}\left(\delta\varphi_z(t)-\left(\tau-\frac{z}{v}\right)\delta\varphi'_z(t)\right)dz\end{aligned}$$

We obtain:

$$\phi_e(t)=\tau_A\int_{L_A}^{L}\delta\varphi'_z(t)\,dz+\int_{0}^{L_A}\frac{z}{v}\delta\varphi'_z(t)\,dz$$

We calculate the autocorrelation function of $\phi_e(t)$ .

$$\begin{aligned}R_e(u)&=\overline{\phi_e(t)\phi_e(t-u)}=\tau_A^2\int_{L_A}^{L}\overline{\delta\varphi'_z(t)\delta\varphi'_z(t-u)}\,\Delta L\,dz\\&\quad+\int_{0}^{L_A}\left(\frac{z}{v}\right)^2\overline{\delta\varphi'_z(t)\delta\varphi'_z(t-u)}\,\Delta L\,dz\end{aligned}$$

where we have again assumed that the fiber noise time derivative is not correlated in position.

We deduce, assuming that $\overline{\delta\varphi'_z(t)\delta\varphi'_z(t-u)}$ is independent of z:

$$R_e(u)=\overline{\delta\varphi'(t)\delta\varphi'(t-u)}\,\Delta L\left(\tau_A^2(L-L_A)+\frac{L_A^3}{3v^2}\right)$$

Finally the phase noise PSD at the output end is:

$$\begin{aligned}S_e(f)&=\left(\frac{L_A}{L}\right)^2\left(1\text{-}\frac{2L_A}{3L}\right)\mathcal{F}\left(\overline{\delta\varphi'(t)\delta\varphi'(t-u)}\,\Delta L\,L\right)\\&=F\,S_0(f)\end{aligned}$$

where $F=\left(\frac{L_A}{L}\right)^2\left(3-2\frac{L_A}{L}\right)$.

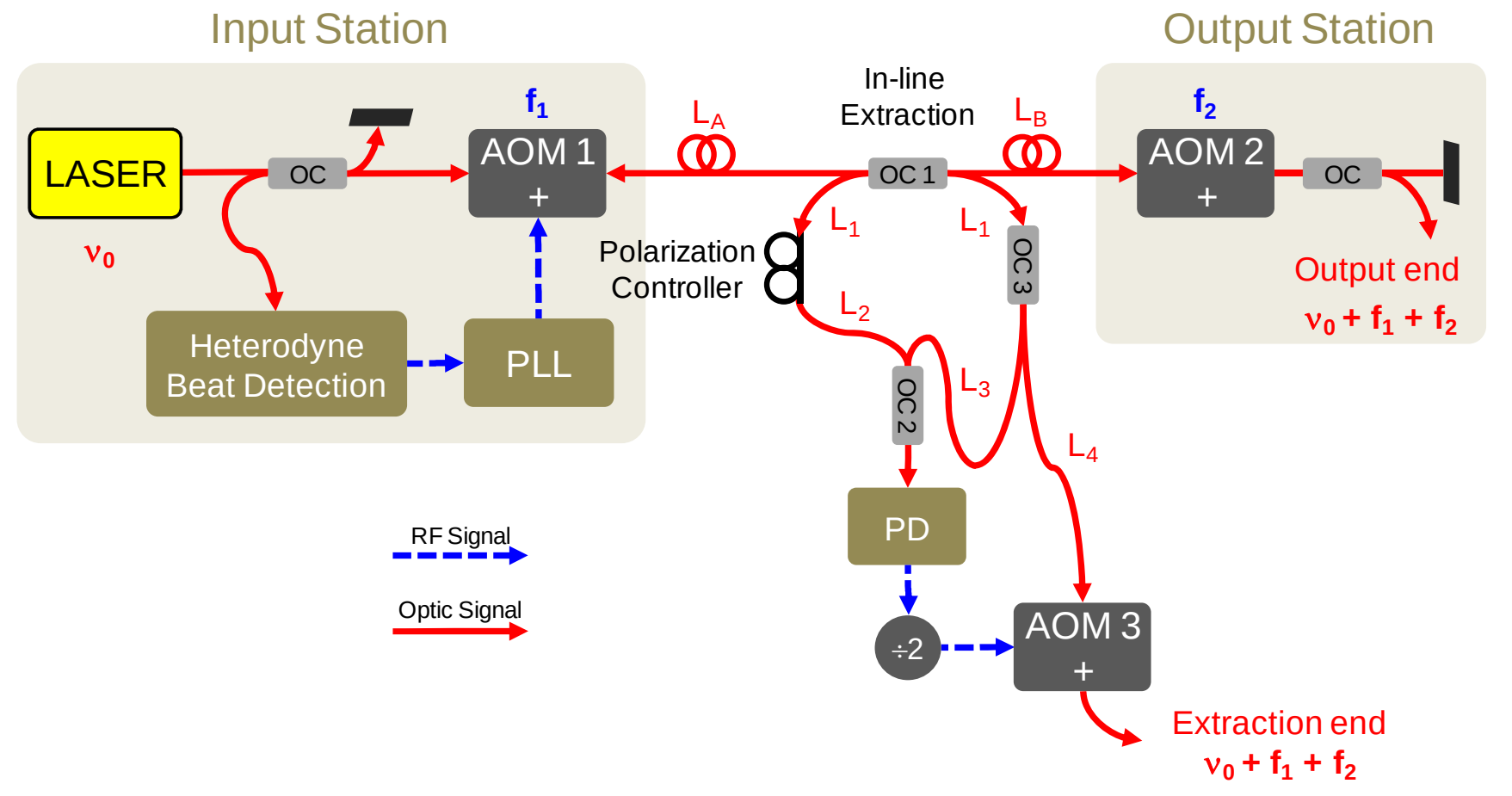

Fig. 1 : Extraction set-up ; AOM: acousto-optic modulator, OC: optical coupler, PD: photodiode, PLL: phase-lock loop

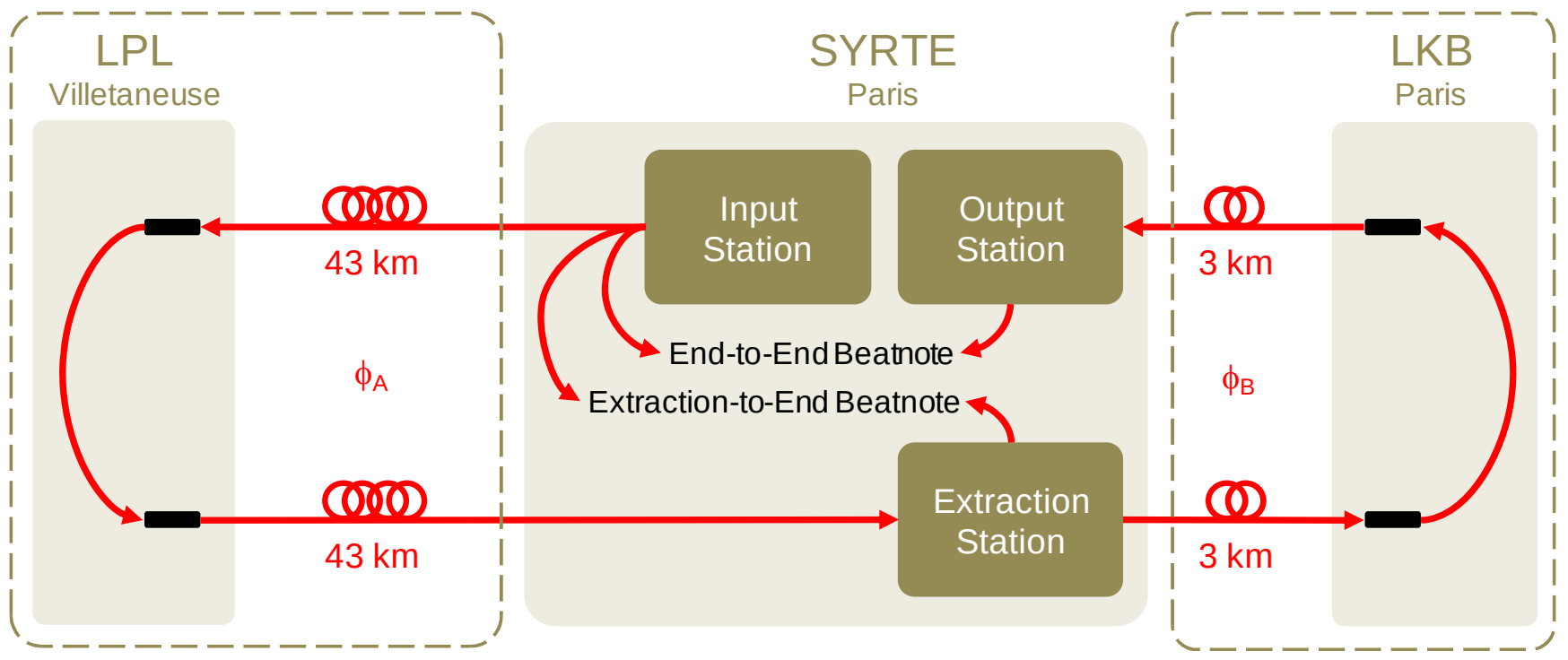

Fig. 2 : Scheme of the urban fiber link composed of two sections of 6-km and 86-km; the left and right sections of fibers (inside the dotted boxes) are exchanged when the extraction is performed near the input end.

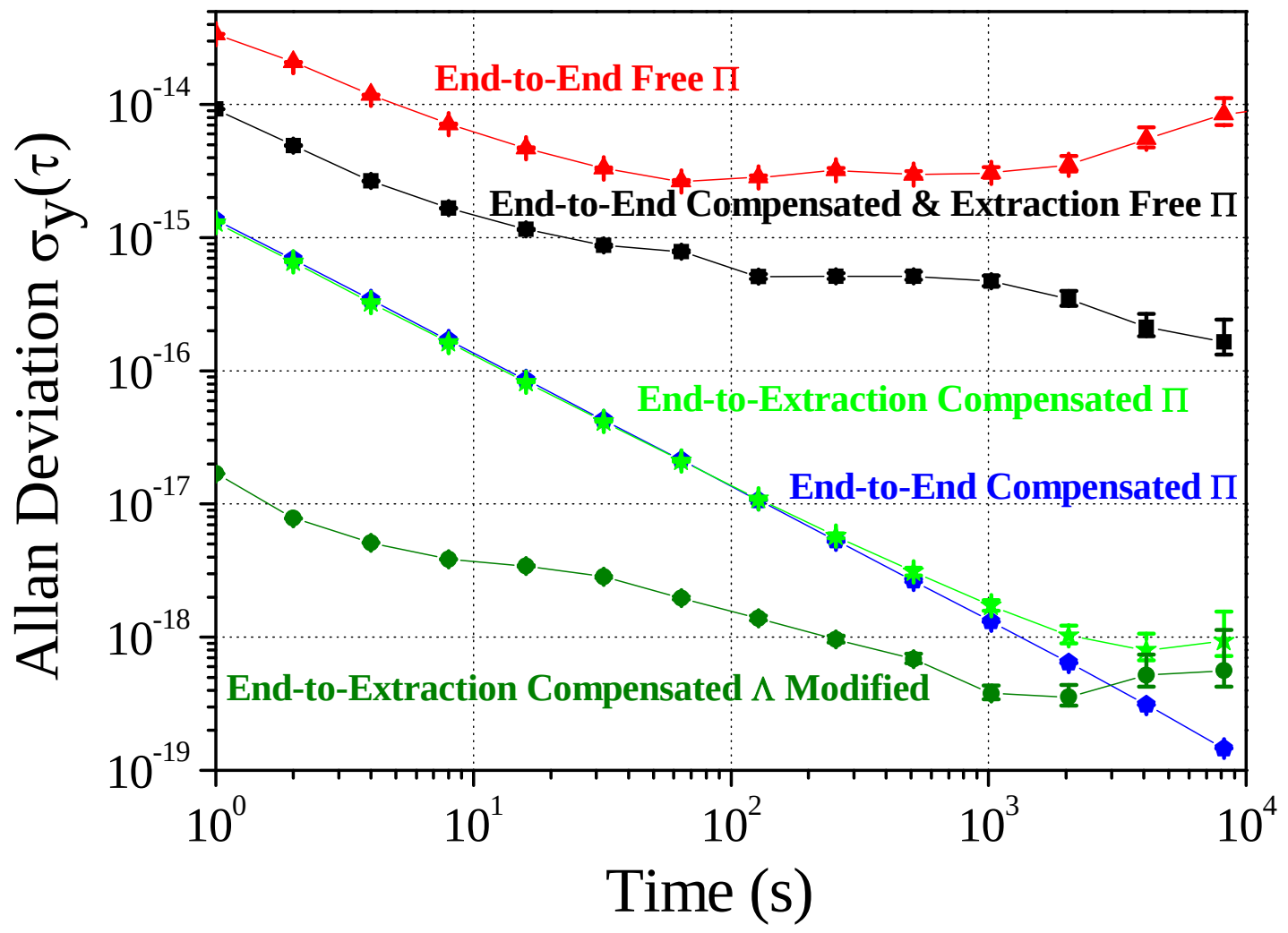

Fig. 3a : Fractional frequency stability when the extraction is done after 86-km of the 92-km main link.: end-to-end free-running main link (red triangles), free-running extraction end (with the main link compensated)(black squares), end-to-end compensated main link (blue diamonds), extraction end (green stars and dark-green circles). All the stabilities are calculated from Π-type samples using overlapping Allan deviation, except the second extraction end stability (dark-green circles), which is calculated from Λ-data using modified Allan deviation, to show the detection floor limitation.

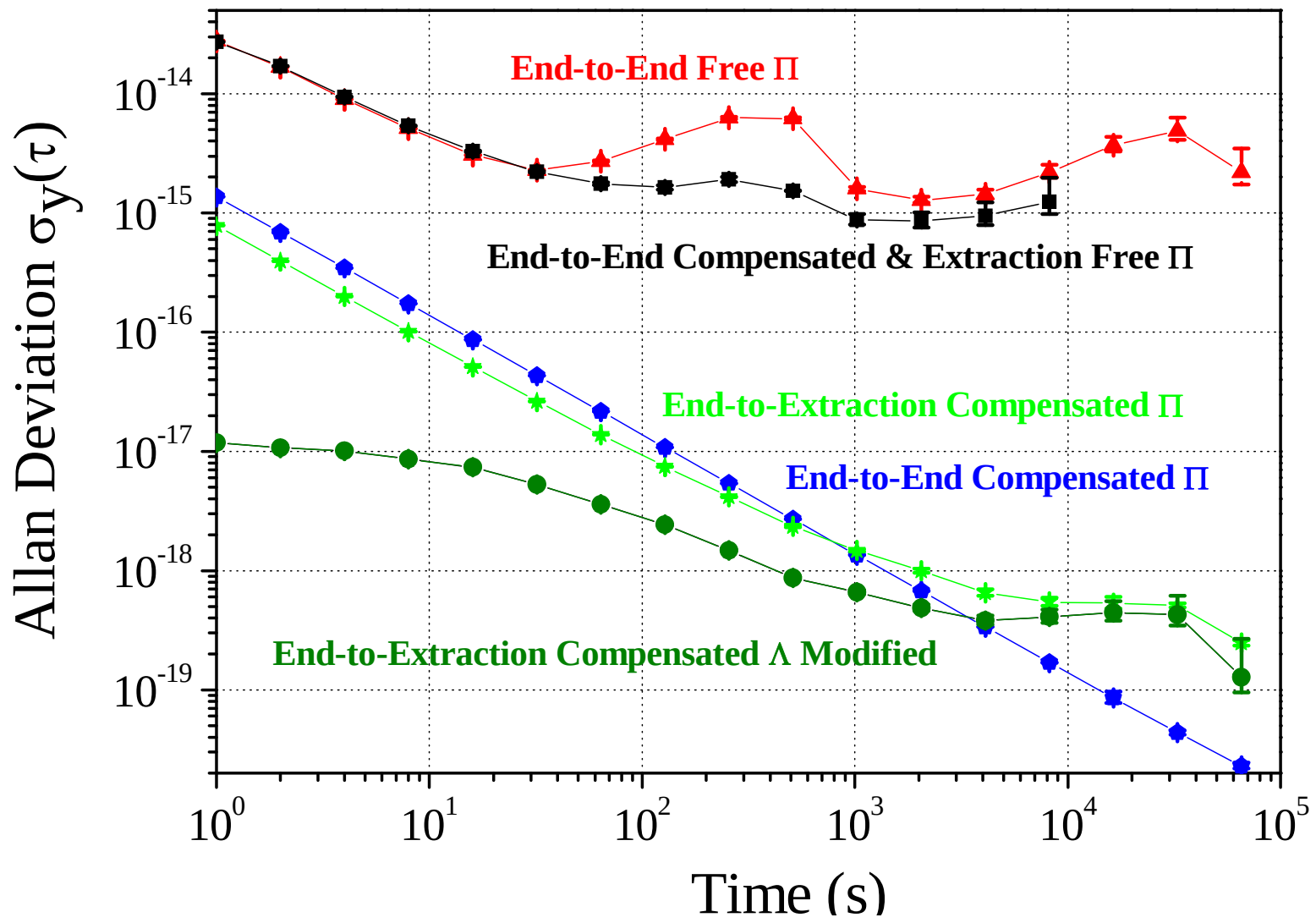

Fig. 3b : Fractional frequency stability when the extraction is done after 6-km of the 92-km main link.: end-to-end free-running main link (red triangles), free-running extraction end (with the main link compensated)(black squares), end-to-end compensated main link (blue diamonds), extraction end (green stars and dark-green circles). All the stabilities are calculated from Π-type samples using overlapping Allan deviation, except the second extraction end stability (dark-green circles), which is calculated from Λ-data using modified Allan deviation, to show the detection floor limitation.

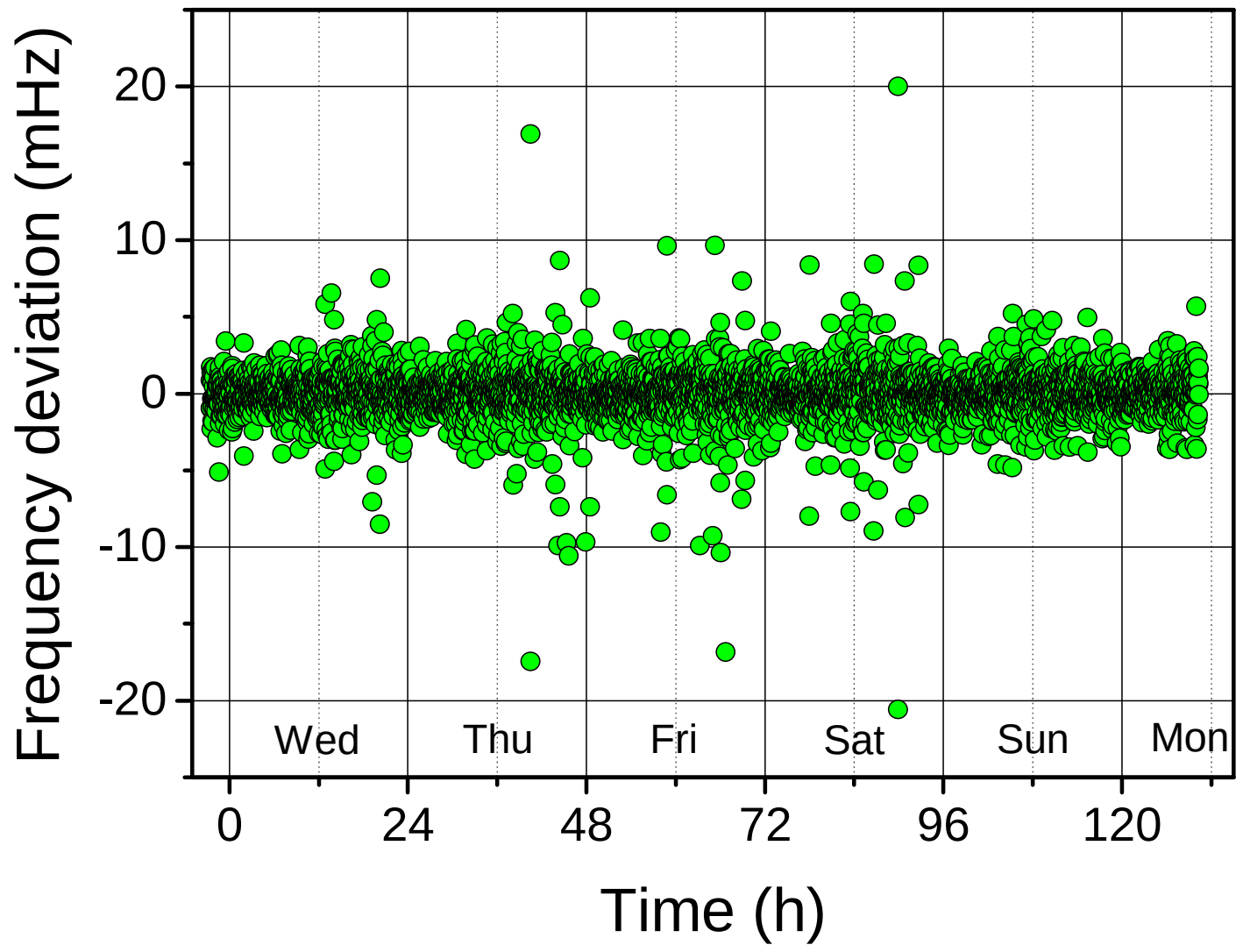

Fig. 4 : Frequency comparison between input and extracted frequency for 5.5 days. Each point corresponds to the mean frequency over 100 data measured with a dead-time free Π-type counter with 1 s gate time.

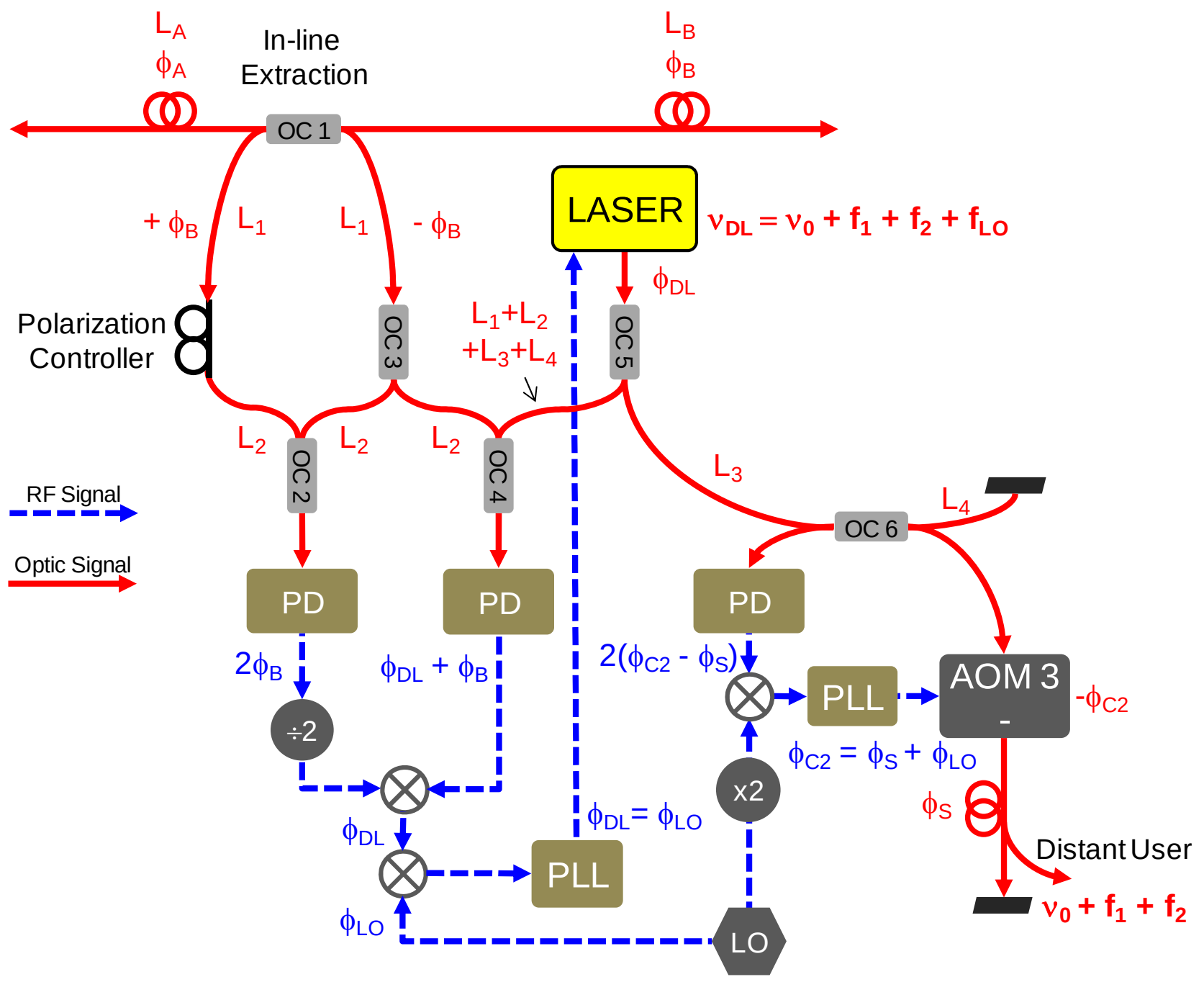

Fig. 5 : Improved extraction set-up with secondary link ; AOM: acousto-optic modulator, OC: optical coupler, PD: photodiode, PLL: phase-lock loop. For clarity, only the phase perturbations (and not the total phase) are indicated.